\documentclass[preprint,aps,showpacs,pre]{revtex4}
\usepackage{amsmath}
\usepackage[dvips]{graphicx}

\begin{document}

\title{Fourth order nematic elasticity and modulated nematic phases: a poor man's approach.}
\author{G. Barbero$^{1}$ and I. Lelidis$^{1,2}$}
\affiliation{$^1$Dipartimento di Scienza Applicata del Politecnico di Torino,\\
Corso Duca degli Abruzzi 24, 10129 Torino, Italia.\\
$^2$ Faculty of Physics, National and Kapodistrian University of Athens, \\
Panepistimiopolis, 15784 Zografos, Athens, Greece}
\date{\today}

\begin{abstract}
We propose an extension of Frank-Oseen's elastic energy for bulk nematic liquid crystals which is based on the hypothesis that the fundamental deformations allowed in nematic liquid crystals are splay, twist and bend. The extended elastic energy is a fourth order form in the fundamental deformations. The existence of bulk spontaneous modulated or deformed nematic liquid crystal ground states is investigated. The analysis is limited to bulk nematic liquid crystals in the absence of limiting surfaces and/or external fields. The non deformed ground state is stable only when Frank-Oseen's elastic constants are positive. In case where at least one of them is negative, the ground state becomes deformed. The analysis of the stability of the deformed states in the space of the elastic parameters allows to characterize different types of deformed nematic phases. Some of them are new nematic phases, for instance a twist -- splay nematic phase is predicted. Inequalities between second order elastic constants which govern the stability of the twist--bend and splay--bend state are obtained.

\end{abstract}

 \maketitle

\section{Introduction}

Conventional achiral nematics have positive elastic constants and therefore do not deform spontaneously. In fact, at equilibrium any deviation of the nematic director, $\mathbf{n}$, from a uniform and non--deformed state is attributed either to the boundary conditions at the interfaces and/or in external bulk fields. Chiral nematics present a modulation of the director along one or two axes known as cholesteric nematic and blue phases respectively. In the case of thin films, modulated structures may appear even in achiral media as liquid crystal nematic and smectic thin films \cite{OP,seul,pang,galerne,stripesLB,yodh,cazabat,anatoly}, magnetic films \cite{bogdanov}, etc.
The existence of a nematic phase that spontaneously  deforms in the bulk of an achiral medium was first predicted by Meyer \cite{meyer} and later by Dozov \cite{dozov}, and Memmer \cite{memmer}. Dozov's paper triggered the revival of experimental interest in bulk modulated nematic phases. A modulated nematic bulk phase, known as the twist-bend nematic, N$_{TB}$ phase, was experimentally observed for the first time only recently \cite{1tbobs,panov}. Many other observations of the N$_{TB}$ phase followed \cite{imrie,chen,Borshch,Zimmer,vana_exp} and a few models have been proposed \cite{shamid,virga,greco,pre,longa,kats,matsu,vana_th} to describe the new nematic ground state. In N$_{TB}$ phase the director exhibits periodic twist and bend deformations forming a conical helix with doubly degenerate domains of opposite handedness. Another modulated nematic phase predicted in \cite{dozov} is the splay--bend nematic, $N_{SB}$, phase but never experimentally observed.

In the analysis presented in \cite{dozov}, the existence of a twist-bend and a splay-bend nematic phase was related to a negative elastic constant of bend. The analysis of the pattern in the ground state was done by means of an elastic energy density containing a term quadratic in the second order derivative of the nematic director, according to a rule proposed some time ago to describe the surface deformation related to the splay-bend elastic constant $K_{13}$ \cite{oldano}.
In a recent paper \cite{lelidis}, we  proposed an elastic energy for the deformed state, in a one dimensional approach, whose deformed state is completely characterized by a tilt angle $\theta$, containing a term proportional to $(\mathrm{d}\theta/\mathrm{d}z)^4$, where $z$ is the coordinate along which $\theta$ is changing. In a more general case where the nematic deformation is characterized by  two angles, varying in the space, the analysis proposed in \cite{lelidis} has to be extended. This is the goal of the present paper. Following the same strategy used in \cite{lelidis}, we generalize the elastic energy density proposed by Oseen, Zocher, and Frank \cite{Oseen,Zocher,Frank}, including fourth order terms in the fundamental deformations characterizing the nematic state. From the stability analysis of the extended up to fourth order energy density, we show that when at least one of the Frank-Oseen's elastic constants is negative then, non--uniform, modulated nematic phases may appear as ground states. A classification of the non-uniform bulk nematic phases is proposed. A phase diagram is calculated for the modulated nematic phase with two elementary nematic deformations in the space of the elastic parameters.

The outline of the paper is as follows. In Section II, a simple extension  of Frank-Oseen's elastic energy is proposed up to the fourth order in the elementary deformations, and the stability conditions for deformed and non-deformed ground states are presented. In Section III, the deformed solutions are classified in respect to the number of elementary deformations. Stability is investigated for each solution, as well as, their relative stability. Some cases of particular interest are further investigated. The paper concludes with a discussion in Section IV.


\section{Elastic energy of a nematic liquid crystal}
Nematic liquid crystals are characterized, from the optical point of view, by an optical axis coinciding with average molecular orientation, which is called nematic director, and usually indicated by ${\bf n}$. The elastic theory for these media is build assuming as deformation tensor the spatial derivatives of the director. The fundamental deformations for a nematic liquid crystal are called splay, twist, and bend \cite{prost,KL}, and are defined by
\begin{equation}
\label{0}s=\nabla \cdot {\bf n},\quad t={\bf n}\cdot (\nabla\times {\bf n})-q_0,\quad{\rm and}\quad {\bf b}={\bf n}\times{\nabla\times {\bf n}},
\end{equation}
where $q_0$ is related to the natural molecular chirality of the molecules forming the nematic phase. Herein, we assume that the molecules are achiral and therefore we take  $q_0=0$. We also do not consider surface like terms as the saddle-splay elastic term. Therefore, in the quadratic approximation, the bulk elastic energy density of the nematic phase is given by
\begin{equation}
\label{1}f_F=\frac{1}{2}\left(K_{11}s^2+K_{22}t^2+K_{33}b^2\right)
\end{equation}
If one or more of the Frank-Oseen's elastic constants becomes negative, the elastic energy density has to be expanded up to the fourth order as follows
\begin{eqnarray}
\label{elf}f&=&\frac{1}{2}\left(K_{11} s^2+K_{22}t^2+K_{33}b^2\right)\nonumber\\
&+&\frac{1}{4}\left(H_{11}s^4+H_{22}t^4+H_{33}b^4+2M_{12}s^2t^2+2M_{13}s^2b^2+2M_{23}t^2b^2\right).
\end{eqnarray}
This is due to the fundamental assumption, valid in elastic theory, according to which in the stable state the elastic deformation has to be finite. From this forth order expansion of the elastic energy one can define the ratio of forth order over second order elastic constants $\sqrt{H/K}$ that has the dimension of a length.  This new length has to be related with the periodicity of the nematic director in non-uniform nematic phases.
The stable states are defined by
\begin{equation}
\label{df}\frac{\partial f}{\partial s}=\frac{\partial f}{\partial t}=\frac{\partial f}{\partial b}=0,
\end{equation}
subjected to the conditions
\begin{equation}
\label{stb}\frac{\partial^2 f}{\partial s^2}\ge 0,
\quad{\begin{bmatrix}
   \frac{\partial^2 f}{\partial s^2} & \frac{\partial^2 f}{\partial s \partial t}  \\
    \frac{\partial^2 f}{\partial t \partial s} &  \frac{\partial^2 f}{\partial t^2}
\end{bmatrix}}\geq 0,
\quad {\begin{bmatrix}
   \frac{\partial^2 f}{\partial s^2} & \frac{\partial^2 f}{\partial s \partial t} & \frac{\partial^2 f}{\partial s \partial b} \\
    \frac{\partial^2 f}{\partial t \partial s} &  \frac{\partial^2 f}{\partial t^2} &  \frac{\partial^2 f}{\partial t \partial b} \\
     \frac{\partial^2 f}{\partial b \partial s} &  \frac{\partial^2 f}{\partial b \partial t} &  \frac{\partial^2 f}{\partial b^2}
\end{bmatrix}}\geq 0
\end{equation}
Equations (\ref{df}) have always the trivial solution
\begin{equation}
\label{4} s=t=|{\bf b}|=0,
\end{equation}
corresponding to the non deformed  (uniform) state. The stability of this state requires that, as it follows from the conditions of stability represented by (\ref{stb}),
\begin{equation}
\label{5}K_{11}\geq 0,\quad\quad K_{11} K_{22}\geq 0,\quad\quad K_{11}K_{22}K_{33}\geq 0.
\end{equation}
As  expected the non deformed state is stable only if the elastic constants of Frank-Oseen are positive. Hereafter, trivial solutions are not discussed.

In case of non--trivial solutions implying at least one negative elastic constant, we first examine the condition on the elastic energy stability.
Since the free energy has to be limited from below the quadratic form
\begin{eqnarray}
q_4=H_{11}s^4+H_{22}t^4+H_{33}b^4+2M_{12}s^2t^2+2M_{13}s^2b^2+2M_{23}t^2b^2>0\,,
\end{eqnarray}
formed by the forth order energy terms has to be positive defined. This is the case when:

\begin{equation}\label{q4in}
H_{jj}>0\quad\&\quad H_{jj}H_{kk}>M_{jk}^2\quad\&\quad det[q_4]>0
\end{equation}
where $j,k=1,2,3$, $j\ne k$, and $det[q_4]$ is the determinant of the quadratic form constructed from $q_4$.

In what follows, we investigate some non--trivial configurations of the nematic director that minimise the elastic energy. We distinguish three cases, namely the simple, double, and triple deformations where one, two and all three nematic elementary deformations appear respectively, in the deformed or modulated nematic state.


\section{non--trivial solutions and stability}

\subsection{Simple deformation}

First, we consider simple deformations  where only one type of the elementary deformations, that is,  either splay, or twist, or bend is permitted. Of course, the case of a simple pure twist is well known \cite{prost,KL} and it will not be discussed further in this paper.

Suppose that $s\ne 0$ and $t=|{\bf b}|=0$, that is, a pure splay case. Then the only non trivial solution that minimise the free energy is $s_0=\pm\sqrt{-K_{11}/H_{11}}$ and its existence implies $K_{11}<0$. The corresponding free energy
\begin{eqnarray}
f_s=f(s\ne 0, t=0, |{\bf b}|=0)=-\dfrac{1}{4}\dfrac{K_{11}^2}{H_{11}}<0\,,
\end{eqnarray}
is negative and therefore lower than the energy of the uniform solution. The configuration of ${\bf n}$ is given from the equations \cite{KL}
\begin{eqnarray}
{\bf div\,n}=s_0\quad\&\quad {\bf curl\, n=0}
\end{eqnarray} Therefore the nematic director can be written as the gradient of a scalar $u$, that is, ${\bf n=\bf grad }u/|{\bf \bf grad }u|$.

In the case of a pure bend deformation, $K_{33}<0$, one finds
\begin{eqnarray}
f_b=f(s= 0, t=0, |{\bf b}|\neq 0)=-\dfrac{1}{4}\dfrac{K_{33}^2}{H_{33}}<0\,,
\end{eqnarray}
that is a spontaneous bend deformation in the bulk with $b_0=\pm\sqrt{-K_{33}/H_{33}}$ if $K_{33}<0$. The conditions on the director ${\bf n}$ are \cite{KL}

\begin{eqnarray}
{\bf div\, n}=0\quad\&\quad {\bf n \wedge curl\, n\neq0 \quad\&\quad n \cdot curl\, n=0}
\end{eqnarray}

Of course, one can not fill the space with a simple splay or bend deformation, that is, defects have to be introduced. That is, opposite to pure twist deformation, pure uniform splay or bend in space is impossible for nematics. Pure spontaneous twist deformation in the frame of the present model, that is, in absence of molecular chirality that would contribute a first order term in $f$, arrives for $K_{22}<0$ with a wavevector proportional to $t_0=\pm\sqrt{-K_{22}/H_{22}}$.


\subsection{Double deformation}
In the following, we examine the case of a double type deformation. From the three different configurations that can be considered combining two elementary elastic deformations, we choose to investigate in details the  splay--bend deformation. The analysis for the other two deformations, that is, twist-bend and twist-splay is similar. For a splay--bend deformation one can consider three cases as follows.

\subsubsection{case: $K_{11}<0\quad\&\quad K_{33}>0$}

Suppose that $s\ne 0$, $|{\bf b}|\ne 0$ and $t=0$. Then, after minimisation of the elastic energy, Eq(\ref{elf}), the following non trivial solutions are found
\begin{eqnarray}\label{sbsol}
s_0^2 =\dfrac{K_{33}M_{13}-K_{11}H_{33}}{H_{11}H_{33}-M_{13}^2}\quad\&\quad b_0^2 =\dfrac{K_{11}M_{13}-K_{33}H_{11}}{H_{11}H_{33}-M_{13}^2}\,,
\end{eqnarray}
while the additional inequalities $K_{33}M_{13}>K_{11}H_{33}$ and $K_{11}M_{13}>K_{33}H_{11}$ have to be verified. If only the elastic constant $K_{11}$ is negative, then the latter conditions implie that $M_{13}$ has to be negative too. Substitution of the solutions (\ref{sbsol}) in $f$ gives for the energy density of the  splay--bend deformation
\begin{eqnarray}\label{fsb}
f_{sb}=f(s_0, t=0, b_0)=-\dfrac{1}{4}\dfrac{K_{11}^2H_{33}+K_{33}^2H_{11}-2K_{11}K_{33}M_{13}}{H_{33}H_{11}-M_{13}^2}<0\,,
\end{eqnarray}
$f_{sb}$ is always negative since $H_{11},H_{33}>0$ and $H_{33}H_{11}>M_{13}^2$ as it follows from (\ref{q4in}).
The Hessian of the energy $f_{sb}$, subjected to the conditions (\ref{q4in}), is always positive for real solutions. One can show that the energy of the double deformation is always lower than the energy of simple splay (or bend) deformation $f_{sb}<f_s$, without additional conditions on the elastic constants

\begin{eqnarray}
f_{sb}-f_s=-\dfrac{1}{4}\dfrac{\left( H_{11}K_{33}-2K_{11}M_{13}\right) ^2}{H_{11}\left(H_{11}H_{33}-M_{13}^2\right) }<0\,.
\end{eqnarray}
that is, $f_{sb}<f_s$, if a splay bend deformation is permitted.

In what follows we combine the above inequalities in order to obtain the range of allowed values for the elastic constants, of the forth order energy expansion, that stabilize modulated phases.
The conditions $K_{33}M_{13}>K_{11}H_{33}$ and $K_{11}M_{13}>K_{33}H_{11}$ using $H_{11}>0$, $H_{11}H_{33}>M_{13}^2$ and supposing $K_{11}<0$ while  $K_{33}>0$ result to the inequalities
\begin{eqnarray}\label{range1a}
0>\dfrac{K_{33}}{K_{11}}H_{11}>M_{13}>-\sqrt{H_{11}H_{33}}\\\label{range1b}
\dfrac{H_{11}}{H_{33}}<\dfrac{K_{11}^2}{K_{33}^2}.
\end{eqnarray}
The second inequality is always valid if the first inequality is verified.
At this point, in order to build the corresponding phase diagram, it is convenient to introduce the following reduced elastic constants
\begin{align}
\dfrac{K_{33}}{K_{11}}=\kappa\;;\quad\dfrac{H_{33}}{H_{11}}=h\;;\quad \dfrac{M_{13}}{H_{11}}=\mu
\end{align} where $h$ is always a positive quantity. Additionaly, we define $\kappa^* = \kappa/\sqrt{h}$ and $\mu^* = \mu /\sqrt{h}$.  Subsequently, inequality (\ref{range1a}) rewrites as
\begin{equation}\label{finalineq}
0>\kappa^* >\mu^* > -1
\end{equation}
These inequalities define the range of the $M_{13}$ coupling elastic constant inside which non uniform nematic solution is thermodynamically stable.

\begin{figure}[h]
\centering
\includegraphics[width=10cm]{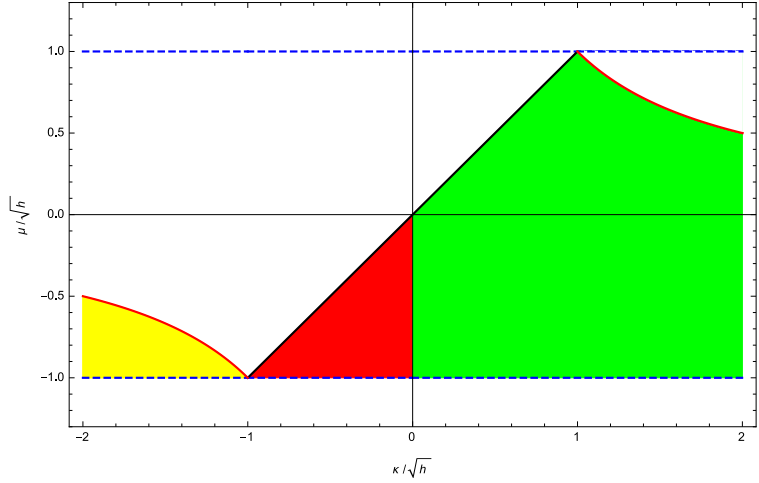}
\caption[]{$(\kappa^*,\mu^*)$ phase diagram for a splay-bend nematic. The Hessian of the free energy is positive between dashed blue horizontal lines. Black diagonal line $\mu^*=\kappa^*$, red lines $\kappa^*\mu^*=1$. The filled regions correspond to minima of the energy. Case: (i) $K_{11}<0$, $K_{33}>0$, red triangular domain, (ii) $K_{11}>0$, $K_{33}<0$ yellow domain, and (iii) $K_{11}<0$, $K_{33}<0$ green domain.}
\label{Figure_1}
\end{figure}

 Figure \ref{Figure_1} gives a graphical representation, in the plane ($\kappa^*,\mu^*$), of the stability domain for non trivial solutions of splay-bend type. Extrema of $f_{sb}$ are located in-between the horizontal
 dashed blue-lines $\mu^{*^2}=1$. The ground states of the splay-bend nematic, described by inequalities (\ref{finalineq}), are localized in the red triangular domain.

\subsubsection{case: $K_{11}>0\quad\&\quad K_{33}<0$}

In a similar way as in the previous case, we find the following inequalities that  have to hold in order to obtain a stable splay--bend nematic phase
 \begin{eqnarray}
 0>\dfrac{K_{11}}{K_{33}}H_{33}>M_{13}>-\sqrt{H_{11}H_{33}}\\
 \dfrac{H_{11}}{H_{33}}>\dfrac{K_{11}^2}{K_{33}^2}.
 \end{eqnarray} or in reduced units
\begin{equation}
0>\dfrac{1}{\kappa^*}>\mu^*>-1
\end{equation}
This type of splay-bend solutions are represented in Figure 1, by the yellow parabolic domain delimitated by the lines $\mu^* =-1$, and $\kappa^*\mu^*=1$.

Note that when one and only one elastic constant is negative then  $M_{13}$ has to be negative too in order to obtain a ground state with a double deformation of the nematic director. For  $M_{13}=0$, solutions (\ref{sbsol}) reduce to $s_0^2 =-K_{11}/H_{11}$ and $b_0^2 =-K_{33}/H_{33}$, that is, both elastic constants have to be negative.

\subsubsection{case: $K_{11}<0\quad\&\quad K_{33}<0$}
Finally, if one considers the case where both,  $K_{11}, K_{33}$, elastic constants are negative, that is $\kappa^*>0$ and in absence of twist deformation, then the domain of $M_{13}$ extends towards positive values too and $1>\mu^*>-1$. Additionally, in order that the splay-bend deformation be stable
the following inequalities have to be verified
\begin{eqnarray}
\kappa^* >\mu^* > -1\quad\text{and}\quad
\dfrac{1}{\kappa^*}>\mu^*>-1.
\end{eqnarray}
The range where splay-bend deformation may exist for both $K_{11}, K_{33}$ negative is shown in Figure 1 (green domain) and it extends also for positive coupling elastic constants.

Considering now the case where bend is zero and looking for double deformation solutions one finds the predicted twist-bend phase and a twist-splay phase (TS). The latter twist-splay phase is a new modulated nematic phase never predicted before. For twist-bend, as well as, twist-splay deformations, the analysis is similar to the splay-bend case treated above and it is not presented here. Therefore, we only give the corresponding solutions that minimise the elastic energy Eqn(\ref{elf}) for the twist-bend

\begin{eqnarray}
b_0^2 =\dfrac{K_{22}M_{23}-K_{33}H_{22}}{H_{22}H_{33}-M_{23}^2}\quad\&\quad t_0^2 =\dfrac{K_{33}M_{23}-K_{22}H_{33}}{H_{22}H_{33}-M_{23}^2}\quad\&\quad s_0=0
\end{eqnarray}
and twist-splay phases.
\begin{eqnarray}
s_0^2 =\dfrac{K_{22}M_{12}-K_{11}H_{22}}{H_{11}H_{22}-M_{12}^2}\quad\&\quad t_0^2 =\dfrac{K_{11}M_{12}-K_{22}H_{11}}{H_{11}H_{22}-M_{12}^2}\quad\&\quad b_0=0
\end{eqnarray}

The phase diagram remain the same as the one shown in Figure 1, but $\kappa^*,\mu^*$  have to be reinterpreted in function of the relevant elastic constants for each case. For instance, the known twist-bend nematic with  $K_{33}<0$, and $K_{22}>0$ corresponds to the yellow domain in Figure 1. The red domain corresponds to a new (of different origin) twist-bend phase which has a negative $K_{22}<0$ while $K_{33}>0$. Finally, a TB-phase with both elastic constants negative is stable in the green domain.

Finally, all the previous stability analysis is based on the hypothesis of a double deformation. The latter is always stable in respect to simple deformations. Nevertheless, if a triple deformation is possible then the above deformations may become metastable states as shown hereafter.



\subsection{Triple deformation: splay-bend-twist}
Suppose now that all three elementary nematic deformations are present, $s\ne 0$, $|{\bf b}|\ne 0$ and $t\ne 0$. Then the non--trivial solutions are

\begin{eqnarray}\label{3s}
s_{\pm}=\pm\,\dfrac{A_{11}}{\Pi}\\
\label{3t}
t_{\pm}=\pm\,\dfrac{A_{22}}{\Pi}\\
\label{3b}
b_{\pm}=\pm\,\dfrac{A_{33}}{\Pi}
\end{eqnarray}
where:

$A_{11}=\sqrt{K_{11}( M_{23}^2-H_{22} H_{33}) + H_{33} K_{22} M_{12} - K_{33} M_{12} M_{23} + H_{22} K_{33} M_{13} - K_{22} M_{13} M_{23}}$

$A_{22}=\sqrt{K_{22}( M_{13}^2-H_{11} H_{33}) + H_{33} K_{11} M_{12} - K_{33} M_{12} M_{13} + H_{11} K_{33} M_{23} - K_{11} M_{13} M_{23}}$

$A_{33}=\sqrt{K_{33}( M_{12}^2-H_{11} H_{22}) + H_{22} K_{11} M_{13} - K_{22} M_{12} M_{13} + H_{11} K_{22} M_{23} - K_{11} M_{12} M_{23}}$

$\Pi=\sqrt{H_{11}H_{22}H_{33}+2M_{12} M_{13} M_{23}-H_{11}M_{23}^2-H_{22}M_{13}^2-H_{33}M_{12}^2}$

The expressions under the square roots have to be positive reals, imposing some conditions to be verified from the elastic constants.

In order to investigate if a non--uniform nematic state with a triple deformation could be a ground state of the system, we have to compare its energy with the energy of a non-deformed nematic. After some algebra, one finds that the energy of the deformed state is given from the expression

\begin{eqnarray}
f_{stb}=-\dfrac{1}{4\Pi^2}\left[  K_{11}^2\left( H_{22} H_{33}-M_{23}^2\right) +K_{22}^2\left( H_{11} H_{33}-M_{13}^2\right) +K_{33}^2\left( H_{11} H_{22}-M_{12}^2\right)\right. \nonumber\\ \left.
+2 K_{11}K_{22}\left(M_{12}H_{33}-M_{13}M_{23} \right)  +2 K_{11}K_{33}\left(M_{13}H_{22}-M_{12}M_{23} \right)\right. \nonumber\\ \left.  +2 K_{22}K_{33}\left(M_{23}H_{11}-M_{12}M_{13} \right) \right]
\end{eqnarray}
This energy is always negative or zero if the solutions given by Eqns (\ref{3s},\ref{3t},\ref{3b}) are real. Therefore, the triple deformed state, if permitted, is always more stable than the uniform one.

In the next paragraph we study the stability of a triple deformation state in respect to the double deformation state.
After some algebra, we find that the energy $f_{sbt}$ of a triple deformation is always lower than the energy $f_{sb}$ of a double deformation since

\begin{eqnarray*}
f_{stb}-f_{sb}=-\dfrac{1}{4}\;\dfrac{\left(K_{22}\left(H_{11} H_{33}-M_{13}^2\right)  -K_{33}\left(M_{23}H_{11}-M_{12}M_{13} \right)-K_{11}\left(M_{13}H_{22}-M_{12}M_{23} \right)\right)^2}{\left(H_{11} H_{33}-M_{13}^2 \right) \Pi^2}<0
\end{eqnarray*}
Similarly one can demonstrate $f_{stb}<f_{12}$ where $1,2=s,t,b$, that is, the triple deformation is the most stable state among the states we investigated, of course under the condition that it is permitted. In the next paragraph, we examine the conditions for the existence of a triple deformation state in some specific cases.

In order to calculate the inequalities to be fulfilled from the elastic constants and consequently to be able to construct a deformation diagram we have to distinguish the following cases: one, two or all three $K_{ii}$ become negatives.



\subsubsection{case: $K_{11}<0\quad\&\quad K_{22}>0\quad\&\quad K_{33}>0$}

Since only $K_{11}<0$, in order that a triple deformation appears, at least the coupling constants $M_{12}$ and $M_{13}$ should be negatives. In addition, the solutions expressed by Eqns (\ref{3s},\ref{3t},\ref{3b}) have to be real. Taking for simplicity $H_{jj}=H$ we find that for $|K_{jj}|=K$, no solution with all three deformations different from zero exists. Such solutions appear only for
\begin{equation}\label{sbst}
|K_{11}|\gtrsim\frac{K_{22}+K_{33}}{2}
\end{equation}
otherwise the splay--bend or the splay-twist deformation is stabilized.
This last inequality for the stabilisation of a twist-bend nematic, when  $K_{11}>0, K_{22}>0, K_{33}<0$, becomes
\begin{equation}\label{tbst}
|K_{33}|\gtrsim\frac{K_{11}+K_{22}}{2}
\end{equation}
Finally, for $K_{11}>0, K_{22}<0, K_{33}>0$, one finds a similar inequality
\begin{equation}\label{}
	|K_{22}|\gtrsim\frac{K_{11}+K_{33}}{2}
\end{equation} for the stabilisation of the TB or TS nematic.
Inequalities (\ref{sbst},\ref{tbst}) are of some interest with respect to the observed twist-bend and splay-bend nematics phases. The twist-bend phase has been observed and its elastic constants have been measured in a few liquid crystals \cite{adlem,yun}. Using these values of the elastic constants, we find that the inequality (\ref{tbst})
does not hold and therefore the observed double deformation of the twist--bend nematic is supported by the present model in respect to the triple deformation. The latter, always is the ground state if it exists, as discussed previously. When inequality (\ref{sbst})  ((\ref{tbst})) holds then the splay--bend (twist--bend) nematic phase is destabilized and a triple deformed nematic state appears.

\subsubsection{case: $K_{11}<0\quad\&\quad K_{22}<0\quad\&\quad K_{33}>0$}

Supposing $H_{11}=H_{22}=H_{33}=H>0$, and taking $K_{11}=K_{22}=-K$ and $K_{33}=K$, one finds that a triple deformation could exist if at least two coupling constants $M_{ij}$ are negative. Of course the one elastic constant approximation is a special case that doesn't appear in real nematics. Therefore we have investigated also the case where $K_{33}>>K$. We found that a triple deformation is still permitted if at least one coupling constant involving bend is negative and in addition the coupling constant ($M_{12}$) between the elementary deformations with negative $K_{jj}$ is negative too.
\subsubsection{case: $K_{11}<0\quad\&\quad K_{22}<0\quad\&\quad K_{33}<0$}

When all three second order elastic constants are negative,  $K_{jj}<0$, then taking, for simplicity, all the coupling elastic constants, $M_{ij}=0$ one finds that
Eqns (\ref{3s},\ref{3t},\ref{3b}) give real solutions for splay, twist, and bend. Therefore  a triple deformation is permitted in the frame of the present model. Eqns (\ref{3s},\ref{3t},\ref{3b}) give
\begin{equation}
s_0=\pm\sqrt{-\dfrac{K_{11}}{H}},\quad t_0=\pm\sqrt{-\dfrac{K_{22}}{H}},\quad
b_0=\pm\sqrt{-\dfrac{K_{33}}{H}}
\end{equation}
 We note that the inequalities for the existence and stability of a triple deformed nematic state without any assumptions on the values of the Frank-Oseen's elastic constants have been evaluated but their expressions are not really convenient for conclusions and therefore they are not presented here. We do not investigate further the case of triple deformation because there is no experimental evidence of such a new nematic phase up to now.

\section{conclusion}
We extended the Frank-Oseen's elastic energy up to the forth order in the elementary nematic deformations and in absence of chirality, in order to account for negative second order $K_{jj}$ elastic constants. We found that spontaneous elastic deformations arise if at least one of the $K_{jj}$ becomes negative. Seven different deformation ground states have been found including the already known twisted nematic, splay--bend and twist--bend phases. The stability of these phases has been investigated. We found that among deformation states that are allowed to exist the most stable is the one involving the higher number of elementary nematic deformations. In particular, some inequalities among the Frank-Oseen's elastic constants $K_{jj}$ were obtained for the stability of the splay--bend, twist--bend, and twist--splay nematic in respect to the triple deformation nematic phase. Our present approach is valid for deformations which depend on one space direction. In the general case where the deformation depends on more than one space direction, one expects more than six elastic terms of forth order.



\end{document}